\documentclass[twocolumn,twoside,slac_two]{revtex4}
\pdfoutput=1
\usepackage{graphicx}
\usepackage{fancyhdr}
\usepackage{graphics}
\usepackage{epstopdf}
\usepackage{textpos}
\usepackage{xspace}
\usepackage[squaren]{SIunits}
\usepackage{longtable}
\usepackage{upgreek}
\usepackage{amssymb, amsfonts, amsmath}
\usepackage[english]{babel}
\usepackage{hyperref}
\providecommand{\abs}[1]{\lvert#1\rvert}

\newcommand{\MET}{\ensuremath{E_{\textnormal{T}}^{\textnormal{miss}}}\xspace}
\newcommand{\sumpt}{\ensuremath{\sum p_{\textnormal{T}}}\xspace}
\newcommand{\minv}{\ensuremath{M_{\textnormal{(T)}}}\xspace}
\newcommand{\pt}{\ensuremath{p_{\textnormal{T}}}\xspace}
\newcommand{\GeV}{\giga\electronvolt\xspace}
\newcommand{\ttbar}{\ensuremath{\text{t}\overline{\text{t}}}\xspace}

\begin{document}

\title{\centering Model Unspecific Search for New Physics in CMS with 2010 LHC Data}


\author{
\centering
\begin{center}
Paul Papacz on behalf of the CMS Collaboration
\end{center}}
\affiliation{\centering III.~Physikalisches Institut~A, RWTH~Aachen, Templergraben~55, 52062~Aachen, Germany}
\begin{abstract}

We present the results of the Model Unspecific Search in CMS (MUSiC), which
systematically scans the data taken by the CMS detector for deviations from the
Standard Model predictions. Due to the minimal theoretical bias this approach is
sensitive to a variety of models for new physics. Events containing at least one
electron or muon are classified according to their content of reconstructed
objects (muons, electrons, photons, jets and missing transverse energy). A broad
scan of three kinematic distributions in those classes is performed by
identifying deviations from Standard Model expectations, accounting for
systematic uncertainties.

In this particular search data taken by CMS in the
year 2010, corresponding to an integrated luminosity of
\unit{36.1}{\pico\barn^{-1}}, have been
analysed.
%
%
%
\end{abstract}

\maketitle
\thispagestyle{fancy}


\section{Introduction}
The start-up of the LHC brings forth a new era of high energy particle physics.
What we will find is yet unknown, however, there is a large number of theories
predicting possible outcomes. Many of those theories are tested by dedicated
analyses at CMS and the other LHC experiments. However, new physics could as
well manifest itself in ways no-one has thought of yet. For this purpose a
\textit{Model Unspecific Search in CMS}: ``MUSiC'' has been implemented.

MUSiC automatically scans and compares the measured data to the simulated
SM expectations without assuming any specific model of new
physics. Thus, covering a large phase space, it is sensitive to a lot of potential
signals.

Any uncovered significant deviation needs additional interpretation, such that
its origin can be determined. Possible causes could be insufficient
understanding of the collision, event generation or detector simulation, or
indeed genuine new physics in real data. Thus, the output of MUSiC must be seen
as only the first, but important step in the potential discovery of new physics.

More details on this and following topics can be found in the Physics Analysis Summary~\cite{bib:pas}.

\section{Implementation}
New physics usually shows up in distinctive final states and so the first task
is to sort the events.  We consider the following physics objects:
\textit{muons} ($\mu$), \textit{electrons} (e), \textit{photons} ($\gamma$),
\textit{particle flow jets} (anti-$k_{\textnormal{T}}$), and \textit{missing transverse energy} (\MET).

Each event is sorted into precisely one \textit{event class}, which represent a single final
state, depending on its content of reconstructed objects.
Together with the requirement of at least one charged electron or muon in any analysed
event, this leads to about 250 event classes containing at least one event in
data or simulation.

Three kinematic distributions are examined, which are promising to spot new physics:
\begin{itemize}
   \item Scalar sum of the transverse momentum of all participating objects:~\sumpt.
   \item Combined (transverse) invariant mass \minv of the objects in the event class.
   \item \MET in classes containing \MET above our predefined threshold.
\end{itemize}
Out of these distributions \sumpt is the most general observable, sensitive to many new physics models involving heavy new particles or modified high-energy behaviour.
The invariant mass allows the discovery of new resonances.
New physics with heavy or highly boosted ``invisible'' particles will show up in the~\MET distribution.
The kinematic variables are calculated from the objects passing the selection criteria listed below.
Objects not fulfilling those criteria are not considered.

\subsection{Event and Object Selection}
Events are selected by single lepton triggers (muon or electron).
The triggers require a minimal~\pt and additional quality criteria.
Due to the rapidly changing instantaneous luminosity, the thresholds have been increased over time, the highest being \unit{15}{\GeV} for muons and \unit{22}{\GeV} for electrons.
In order to remain well above the trigger thresholds, at least one muon with \unit{25}{\GeV} or one electron with \unit{30}{\GeV} is required.
The trigger efficiency on selected leptons is above \unit{95}{\%}. 

A summary of the selection criteria applied on physics objects in each event is given in
tab.~\ref{tab:criteria}.
%
   \begin{longtable}{l|c|c|l}
   \caption{Selection criteria applied to the physics objects in each event.\label{tab:criteria}}\\
      Object     & $p^{\text{min}}_{\text{T}}$ & $\abs{\eta}$ &                                       Other Cuts \\
      \hline
      $\upmu$    & \unit{18}{\GeV}             & $< 2.1$      &                            isolation/track       \\
      e          & \unit{25}{\GeV}             & $< 2.5$      &                            shape/track/isolation \\
      $\upgamma$ & \unit{25}{\GeV}             & $< 1.4$      &                            shape/isolation       \\
      jet        & \unit{50}{\GeV}             & $< 2.5$      &                            energy fraction       \\
      \MET       & \unit{30}{\GeV}             & $-$          &                                                  \\
   \end{longtable}
%
\vspace{10mm}
To avoid using the same energy entry more than once,
objects are removed if they are too close to each other ($\Updelta R < 0.2$)\footnote{$\Updelta R = \sqrt{\Updelta \phi^2 + \Updelta \eta^2}$}:
Jets are removed if there are photons or electrons nearby, and photons are removed if an electron is close.

\section{Search Algorithm}
\label{sec:algo}
All kinematic distributions are fed through a scanning algorithm that systematically scans for deviations, comparing the simulated SM prediction with the measured data.
Since all distributions are analysed as binned histograms, the bin width is adjusted to the resolution of the considered variable.

For each region of adjacent bins the algorithm calculates the number of expected
events ($B$), its uncertainty ($\sigma$), and the number of observed events ($N$).
The uncertainty calculation takes into account possible correlations between bins
and individual uncertainties (see section~\ref{sec:sys} for more details).
A Poisson tail probability~$p$ can be computed which denotes the probability of a
random fluctuation to be at least as extreme as the observed value.
To incorporate the systematic uncertainties on this probability,
the Poisson distribution is convoluted with a normal distribution:
\begin{align}
   p =
      \begin{cases}
         \displaystyle
         A \cdot \sum\limits^{B}_{i = 0} \int\limits_0^\infty e^{\frac{-(\mu-B)^2}{2\sigma^2}} \cdot \frac{e^{-\mu} \mu^i}{i!}\,\mathrm{d}\mu &\text{if } N < B\\
         \displaystyle
         A \cdot \sum\limits^{\infty}_{i = N} \int\limits_0^\infty e^{\frac{-(\mu-B)^2}{2\sigma^2}} \cdot \frac{e^{-\mu} \mu^i}{i!}\,\mathrm{d}\mu &\text{if } N \geq B
      \end{cases}
\end{align}
The factor~$A$ ensures normalisation, since the normal distribution is truncated at~0.
In each distribution, the region with the smallest value of $p =
p_{\textnormal{data}}$ is chosen as the \textit{Region of Interest}~(RoI).

\section{Look-Elsewhere-Effect}
While $p_{\textnormal{data}}$ denotes the probability of each RoI seen individually, it cannot be
used as a statistical estimator for the global significance for such a deviation in any region.
A penalty factor needs to be included to account for the number of investigated regions.
Using $\mathcal{O}(10^5)$ pseudo-experiments, we can compute a new probability $\tilde{p}$ of
measuring at least one deviation in any region in a given distribution with a
lower $p$-value than the one seen in the RoI of the data.
Each pseudo-experiment consists of one pseudo-data histogram for each distribution,
thus representing one possible outcome of a measurement.
The pseudo-data distribution for each class is fed into the same scanning algorithm as described above,
resulting in a collection of p-values $p_{\textnormal{pseudo}}$.
The value $\tilde{p}$ for a given distribution is then simply the number of
pseudo-experiments with $p_{\textnormal{pseudo}} < p_{\textnormal{data}}$, divided by the total number of pseudo-experiments:
\begin{align}
  \tilde{p} = \frac{N_{\textnormal{pseudo}}(p_{\textnormal{pseudo}} < p_{\textnormal{data}})}{N_{\textnormal{pseudo}}}
\end{align}


\section{Systematic Uncertainties}
\label{sec:sys}
This kind of search is highly affected by systematic uncertainties. An overview
of the uncertainties used in this analysis is shown in tab.~\ref{tab:unc}.
Some of the included uncertainties can be correlated between bins of one distribution or across classes.
%
   \begin{longtable}{l|c|c}
   \caption{Overview of the systematic uncertainties used in this analysis.\label{tab:unc}}\\
      Contribution               & Value         & Remarks                                                    \\
      \hline
      MC statistics              & various       & sample dependent                                           \\
      Luminosity                 & \unit{4}{\%}  & \\
      parton density fkt.        & various       & PDF reweighting method                                     \\
      jet energy corr.           & 3~to~5~\%     & $p^{\vphantom{i}}_{\text{\tiny T}}$ and $\eta$ dependent \\
      reconstruction eff.        & 1~to~4~\%     & object dependent                                           \\
      misreconstrction prob.     & 30~to~100~\%  & object dependent                                           \\
      W-boson cross sec.         & \unit{5}{\%}  & NNLO                                                       \\
      Drell-Yan cross sec.       & \unit{5}{\%}  & NNLO                                                       \\
      \ttbar cross sec. 		 & \unit{10}{\%} & NNLL                                                       \\
      di-boson cross sec.        & \unit{10}{\%} & LO/NLO                                                     \\
      $\Upsilon$ cross sec.      & \unit{30}{\%} & measured cross sections                                    \\
      QCD-multijet cross sec.    & \unit{50}{\%} & LO                                                         \\
      photon+jets cross sec.     & \unit{50}{\%} & LO                                                         \\
   \end{longtable}
%
\vspace{5mm}
The reconstruction efficiencies and the misidentification probability are determined from simulation and uncertainties are applied to cover possible differences to data.

\section{Results}
\begin{figure}[htb]
   \centering
   \includegraphics[width=.4\textwidth]{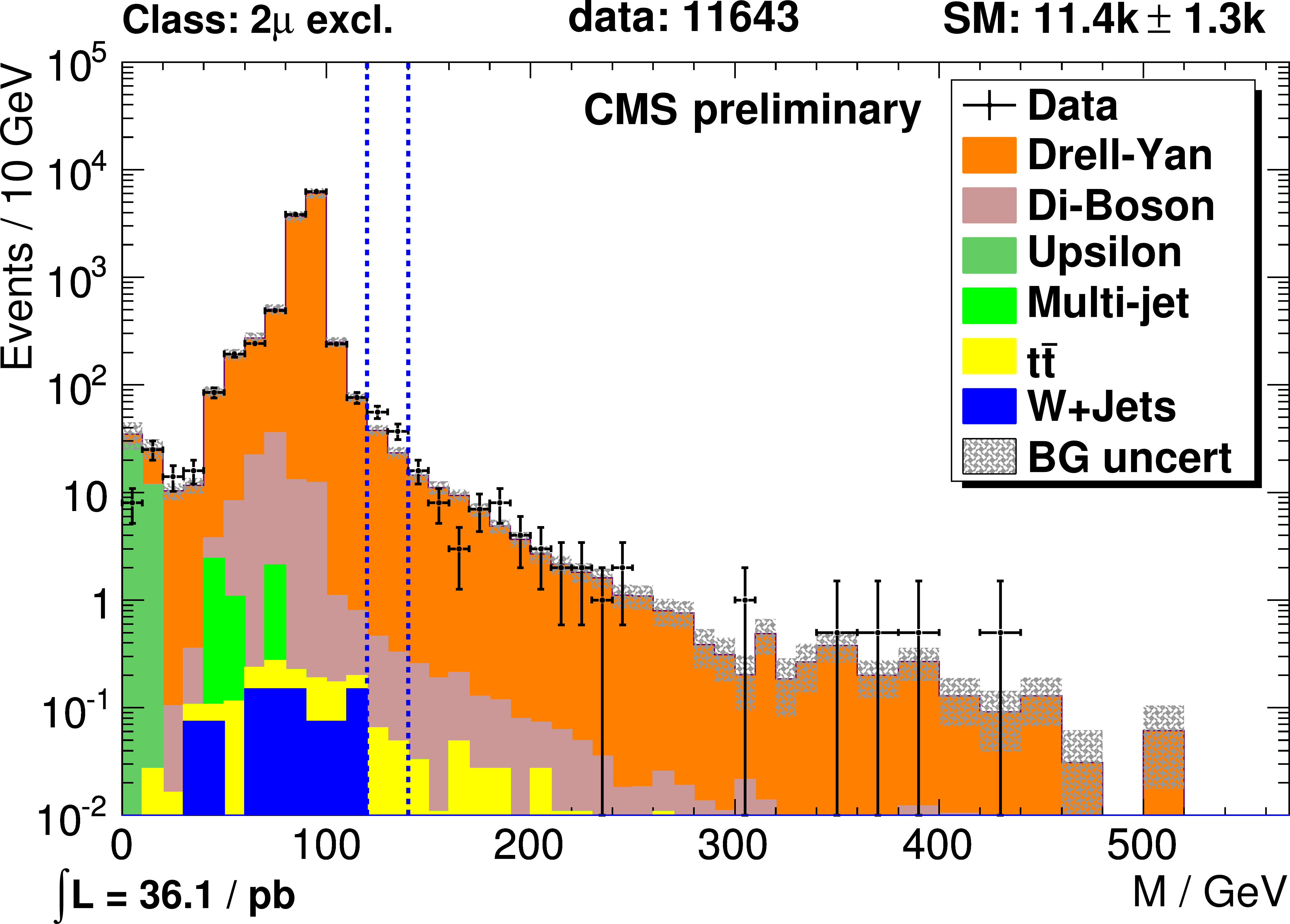}%
   \caption{A typical distribution as seen by the MUSiC analysis. The region in
   between the two blue dashed lines is the Region of Interest. The kinematic
   variable investigated here is the invariant mass of the two muons.}
   \label{fig:2mu}
\end{figure}

\begin{figure}[htb]
   \centering
   \includegraphics[width=.4\textwidth]{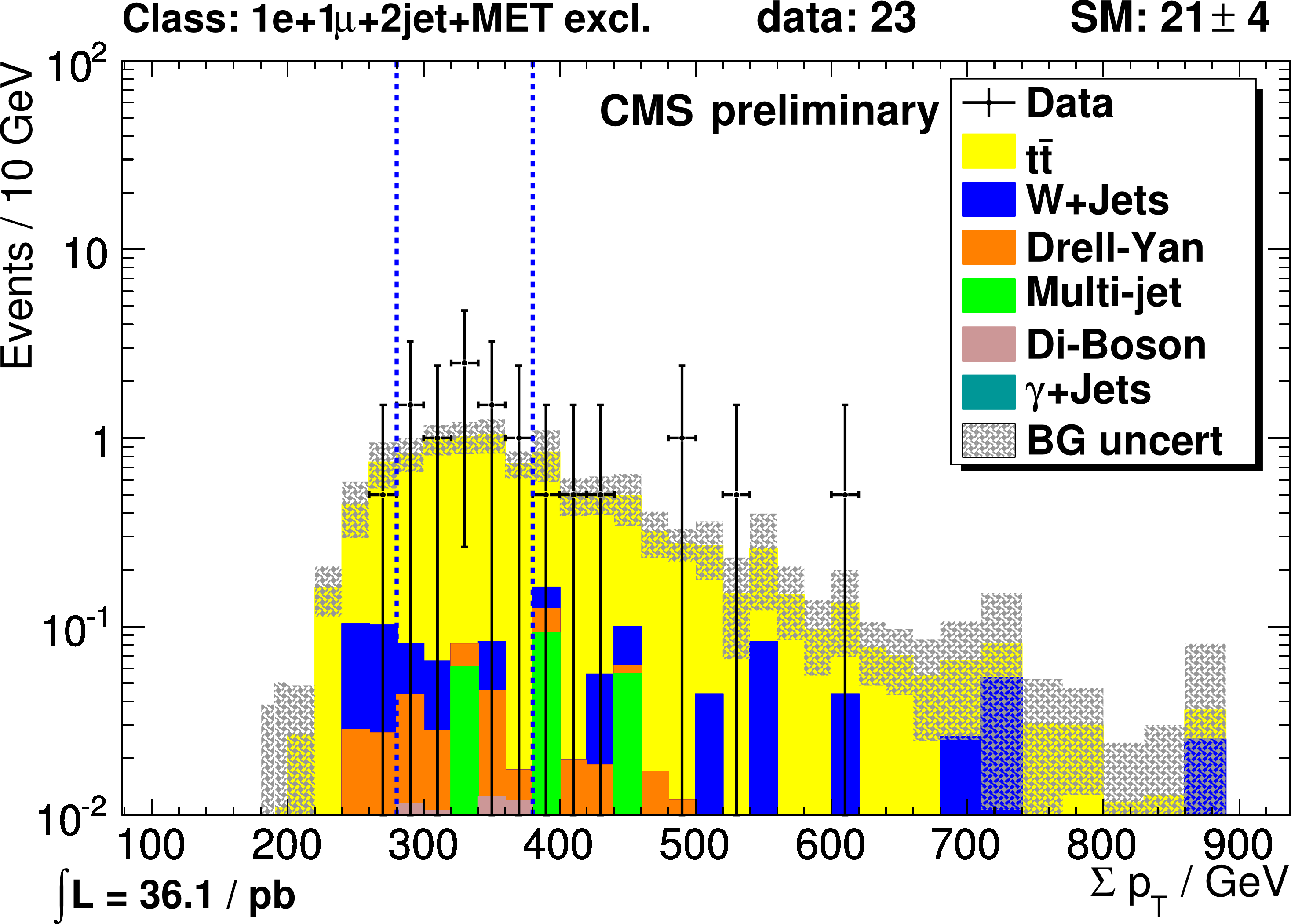}%
   \caption{A typical \ttbar dominated event class with one electron, one muon two
   jets and missing transverse energy in the final state. The kinematic
   variable investigated here is the scalar sum of the transverse momentum
   of all particles in the final state.}
   \label{fig:1e1mu2jet1met}
\end{figure}

As a demonstration two typical distribution are shown. The first one shows a
Drell-Yan dominated class with two muons in the final state (fig.~\ref{fig:2mu})
the second one shows a \ttbar dominated class with one electron, one muon, two
jets and missing transverse energy in the final state~(fig.~\ref{fig:1e1mu2jet1met}).

Overall 287 distribution in 118 event classes have been analysed in 2010 data.

The $\tilde{p}$ distributions of the analysed event classes are determined
separately for the three different kinematic variables.

\begin{figure}[htb]
   \includegraphics[width=.4\textwidth]{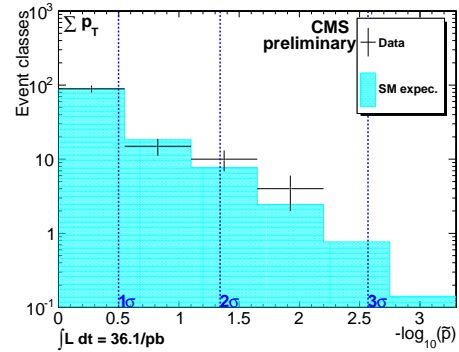}%
   \caption{Distribution of expected (light blue) and observed (crosses)
   $\tilde{p}$ values for 2010 data for \sumpt.}
   \label{fig:p-sum}
\end{figure}

\begin{figure}[htb]
   \includegraphics[width=.4\textwidth]{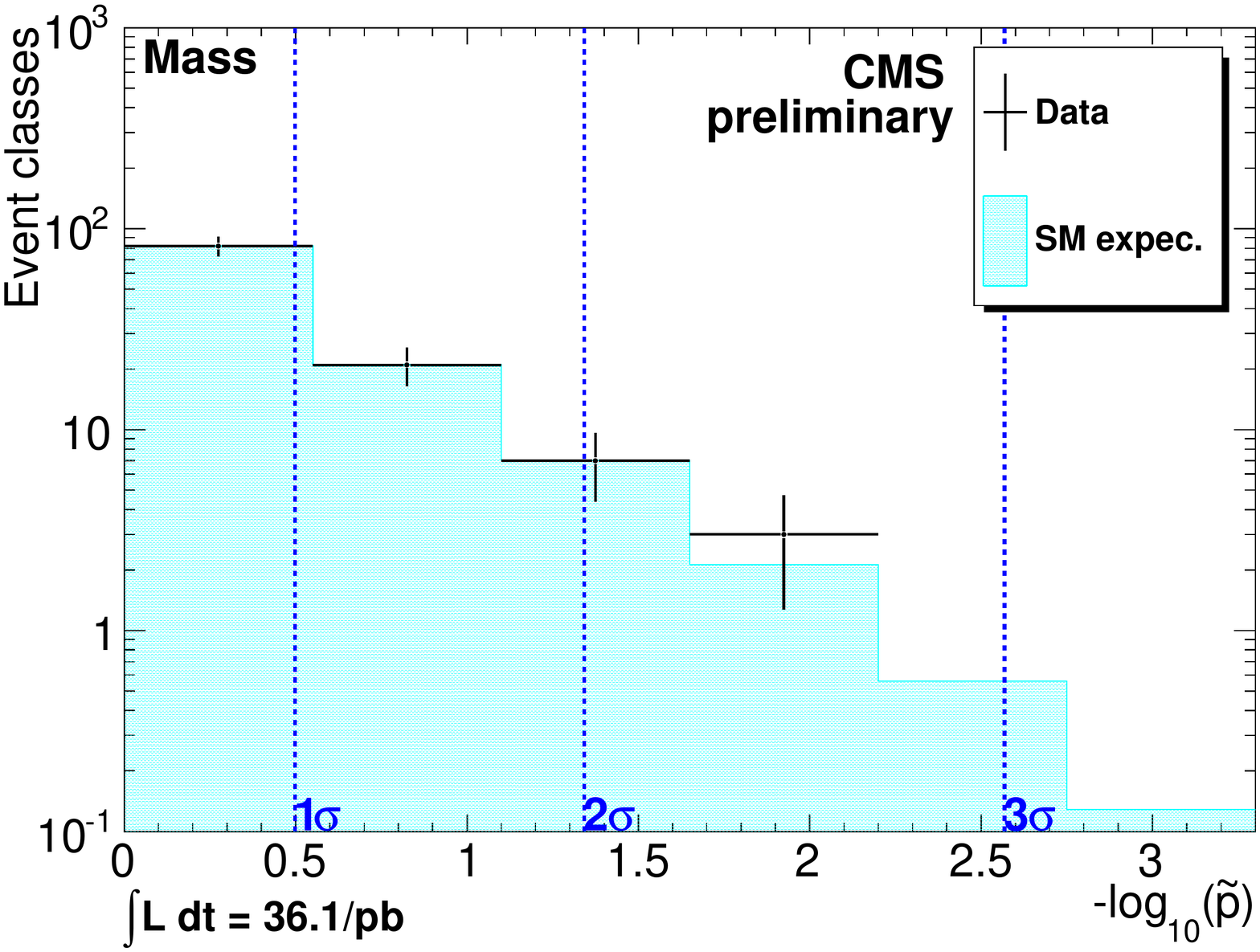}%
   \caption{Distribution of expected (light blue) and observed (crosses)
   $\tilde{p}$ values for 2010 data for \minv.}
   \label{fig:p-minv}
\end{figure}

\begin{figure}[htb]
   \includegraphics[width=.4\textwidth]{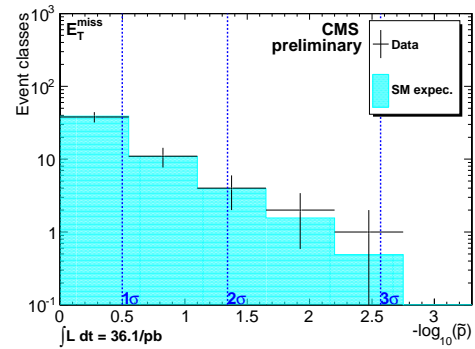}%
   \caption{Distribution of expected (light blue) and observed (crosses)
   $\tilde{p}$ values for 2010 data for \MET.}
   \label{fig:p-met}
\end{figure}
As can be seen in fig.~\ref{fig:p-sum},~\ref{fig:p-minv},~\ref{fig:p-met} the
predictions for the $\tilde{p}$ distributions are very accurate. The used SM
samples describe the data very well.

In the scope of this analysis no deviations from the SM have been found in LHC
2010 data.

\bibliography{pic_proceedings}

\end{document}